\documentstyle[preprint,prl,epsfig,aps]{revtex}
\begin{document}
\tightenlines 

\title{{}Low-mass $\mathrm{e}^+\mathrm{e}^-$ pair production
  in 158\thinspace A GeV Pb-Au collisions at the CERN SPS, its dependence 
on multiplicity and transverse momentum
\thanks{Doctoral Thesis of C.~Voigt.}}
       
\author{{}
G.~Agakichiev$^a$,
R.~Baur$^b$,
P.~Braun-Munzinger$^c$,
F.~Ceretto$^d$,
A.~Drees$^b$\footnote{Now at State University New York at Stony Brook},
S.~Esumi$^b$,
U.~Faschingbauer$^{b, d}$,
Z.~Fraenkel$^e$,
Ch.~Fuchs$^d$,
E.~Gatti$^f$,
P.~Gl\"assel$^b$,
C.\thinspace P.~de los Heros$^e$,
P.~Holl$^g$,
Ch.~Jung$^b$,
B.~Lenkeit$^b$,
M.~Messer$^b$,
Y.~Panebrattsev$^a$,
A.~Pfeiffer$^b$,
J.~Rak$^d$,
I.~Ravinovich$^e$,
S.~Razin$^a$,
P.~Rehak$^g$,
M.~Richter$^b$,
M.~Sampietro$^f$,
N.~Saveljic$^a$,
J.~Schukraft$^h$,
S.~Shimansky$^a$,
W.~Seipp$^b$,
E.~Socol$^e$,
H.\thinspace J.~Specht$^b$,
J.~Stachel$^b$,
G.~Tel-Zur$^e$,
I.~Tserruya$^e$,
T.~Ullrich$^b$,
C.~Voigt$^b$,
C.~Weber$^b$,
J.\thinspace P.~Wessels$^b$,
T.~Wienold$^b$,
J.\thinspace P.~Wurm$^d$,
V.~Yurevich$^a$
}

\author{(CERES Collaboration)}

\address{
{$^a$}{ JINR, Dubna, Russia }\\
{$^b$}{ Universit\"at Heidelberg, Germany }\\
{$^c$}{ GSI, Darmstadt, Germany} \\
{$^d$}{ Max-Planck-Institut f\"ur Kernphysik, Heidelberg, Germany }\\
{$^e$}{ Weizmann Institute of Science, Rehovot, Israel }\\
{$^f$}{ Politecnico di Milano, Italy }\\
{$^g$}{ Brookhaven National Laboratory, Upton, USA }\\
{$^h$}{ CERN, Geneva, Switzerland }\\
}

\date{\today}

\maketitle

\begin{abstract}
  We report a measurement of low-mass electron pairs observed in 158
  GeV/nucleon Pb-Au collisions. The pair yield integrated over the range
  of invariant masses 0.2 $\leq$ m $\leq$ 2.0 GeV/c$^2$ is enhanced by a
  factor of 3.5 $\pm$ 0.4 (stat) $\pm$ 0.9 (syst) over the expectation
  from neutral meson decays. As observed previously in S-Au collisions,
  the enhancement is most pronounced in the invariant-mass region
  300-700 MeV/c$^2$. For Pb-Au we find evidence for a strong increase of
  the enhancement with centrality. In addition, we show that the
  enhancement covers a wide range in transverse momentum, but is largest
  at the lowest   observed $p_\perp$. 
\end{abstract}
\newpage  


Three experiments at the CERN Super-Proton Synchrotron (HELIOS-3
\cite{helios3:sw}, NA38 \cite{na38:su}
and CERES/NA45 \cite{ceres:sau}) have observed significantly more 
dileptons from central collisions of 200 GeV/nucleon sulfur beams with
heavy targets than is expected from `conventional sources' 
\cite{ref1,ref2}.  The
enhancement is most conspicuous for electron pairs in the region of
invariant masses from 300 to 700 MeV/c$^2$ measured close to
midrapidity \cite{ceres:sau}. The comparison to expectations from
conventional sources rests on a simultaneous study of neutral meson 
and e$^+$e$^-$ pair production in 450 GeV p--Be and p--Au
\cite{ceres:pbegamma,ceres:pbe} where it was demonstrated that the
measured e$^+$e$^-$ pair mass spectra are very well described in shape as
well as absolute yield by the direct and Dalitz decays of neutral
mesons. For S-Au, the hadronic sources were scaled proportional to
charged-particle multiplicity.  The observations stimulated an intense
discussion of the processes involved in the emission of low-mass
dileptons from dense and hot hadronic matter.  The extra strength
obtained from $\pi\pi$-annihilation and other scattering processes is
not sufficient to explain the dilepton data
\cite{li-ko-brown,cassing,wambach,theorylist}.  Most theoretical
approaches therefore have focused on the way vector mesons may 
change inside the dense and hot medium. Dropping meson masses, derived
from theoretical arguments related to chiral symmetry restoration
\cite{brown-rho,hat-lee} turned out as a suitable ingredient that
creates sufficient strength below the $\rho/\omega$ resonance.
This allowed to describe simultaneously the sulfur data from CERES and
HELIOS-3 \cite{li-ko-brown,cassing}. Presently, more conventional
approaches in which in-medium modifications of the meson properties
are related to hadronic rescattering processes that broaden the strength
rather than shift it downwards also give good agreement with existing
data \cite{wambach}.

In view of the highly complex physics processes in dense hadron matter
it is clear that more and better data are mandatory. In this letter,
we present first results for the 158 GeV/nucleon Pb-Au collisions in
order to (i) demonstrate that the low-mass dielectron enhancement is
also present in the lead data, (ii) provide first evidence for a
statistically significant stronger-than-linear dependence of the pair
yield on charged-particle multiplicity, and (iii) show the dependence
of the pair yield on the transverse momentum $p_\perp$.  Part of the
data were already presented in preliminary form \cite{ceres:qm96}.

CERES is an experiment especially designed for the detection and
measurement of low mass electron positron pairs. Electron pairs are
reconstructed from invariant masses of a few tens of MeV/c$^2$ up to 2
GeV/c$^2$ in the pseudorapidity region $2.1 < \eta < 2.65$.  Originally,
the setup was optimized to investigate p- and S-induced collisions,
providing electron identification and tracking by two ring-imaging
Cherenkov counters (RICH-1, RICH-2). One is situated before and one
after an axially symmetric magnetic field confined to a narrow region
between the two radiators \cite{ceres:bari}.  While the experiment was
used successfully to measure $\mathrm{e}^+\mathrm{e}^-$ production in
p-Be, p-Au \cite{ceres:pbe} and S-Au \cite{ceres:sau} collisions, an
upgrade was necessary to cope with the high-multiplicity environment in
Pb-Au collisions.  With more than 300 charged particles emitted into the
CERES acceptance in a central collision, the pattern recognition based
solely on the RICH detectors breaks down. Therefore, the spectrometer
was supplemented by charged-particle tracking devices, namely a
closely-spaced doublet of silicon drift detectors (SDD) downstream of
the target and before RICH-1, and a multiwire proportional chamber with
pad readout (PadC) downstream of RICH-2. The information from these
detectors combined with the RICH data restores the necessary
reconstruction efficiency. In addition, they provide extra background
rejection and improve the momentum resolution and, hence, the 
invariant-mass resolution of the spectrometer.  The rate capability of the data
acquisition system was improved by a factor of four allowing the
collection of more than 500 events per 5s burst of the SPS. With this
setup 16$\cdot$10$^6$ events from Pb-Au collisions were recorded during
a 10 day running period in the fall of 1995. The events were selected
requiring a minimum amplitude in a scintillator array downstream of the
spectrometer. The trigger covers a large range of charged-particle
multiplicities in order to study the impact parameter dependence.

In the off-line track reconstruction, first the event vertex is
reconstructed from more than a hundred particle trajectories traversing
both SDDs. Independently, Cherenkov ring candidates are reconstructed
according to a pattern recognition algorithm \cite{ceres:bari} in both
RICH detectors.  Each track segment from the vertex to the SDDs is then
extrapolated into \mbox{RICH-1}, where its angular coordinates are
compared to those of the centers of all ring candidates found.
Approximately 10\% of all SDD trajectories match to a ring center within
a 5 mrad window, and are marked as potential electron track segments.
Many of these do not originate from electrons or positrons since a large
fraction of the ring candidates are accidental combinations of photon
hits. These fake tracks are removed in subsequent steps of the analysis.
All segments are further extrapolated through the magnetic field into
the PadC.  Because the momentum and therefore the deflection in the
magnetic field is not known at this stage of the analysis, every hit in
the PadC within a certain area has to be combined with the track segment
constructed so far. The azimuthal size of this area is defined by a cut
on transverse momentum, $p_\perp \geq$ 50 MeV/c, its size in polar angle
$\theta$ by the combined $\pm$ 3 $\sigma$-widths of detector
resolutions, multiple scattering and higher order effects, which
amounted to $\approx$ 1.8 mrad.  Within these boundaries more than two
hits are found on average, but only the small fraction which matches to
a Cherenkov ring in RICH-2 is accepted. Finally, cuts on the quality of
Cherenkov rings as well as on the track match between individual
detectors reduce the amount of fake tracks to a level below 1\%.
Possible ambiguities in the tracking are removed by requiring the best
match in radial direction between all the detectors.  This scheme leads
to the reconstruction of unique e$^+$e$^-$ tracks, with one exception:
electron and positron tracks originating from photon conversions or
Dalitz pairs of small opening angle are allowed to share the same ring
in RICH-1 since the double-ring resolution in RICH-1 is insufficient to
resolve all of these so-called V-tracks. The momentum is determined
combining all measurements along the track, three space points and two
angles, each weighted according to detector resolution and multiple
scattering. The relative momentum resolution obtained is $(\sigma_p/p)^2
= 0.0012 + 0.001 p^2$ ($p$ in GeV/c), better by
nearly 50\% at high momenta compared to the original setup. \\

Most tracks belong to pairs from photon conversions and $\pi^0$ Dalitz
decays, characterized by small opening angles (close pairs) and low
invariant  masses.  In order to reduce for each track the number of
possible combinations, tracks from unambiguously identified close pairs
are taken out of the sample.  V-tracks are removed.  About 90\% of
all pairs with opening angles smaller than 50 mrad and invariant
masses below 150 MeV/c$^2$ result from Dalitz decays. They are kept in
the sample but marked as assigned tracks, and hence not combined with
others. For a significant fraction of close pairs only one track is
reconstructed due to the finite reconstruction efficiency and acceptance.
Its combination with other tracks creates the  combinatorial pair
background. 

The central problem of the analysis is to recognize and reject as many
of the partially reconstructed conversions and Dalitz decays as
possible.  An important kinematic characteristic of conversions and
Dalitz decays is the softness of the $p_\perp$ spectra of electrons and
positrons compared to those for pairs with masses above 0.2 GeV/c$^2$.
This allows for a drastic suppression of background tracks by requiring
$p_\perp \geq$ 200 MeV/c for each track. Another characteristic is the
small opening angle, and since there is no magnetic field upstream of
and within RICH-1, the opening angle of a pair remains unchanged through
half of the setup.  Thus a pair with small opening angle can be
identified even if the trajectory of one of its tracks is incompletely
measured.  Depending on the size of the pair opening angle, different
information is exploited. (i) Opening angles smaller than the double hit
resolution of the SDD system ($<5$ mrad) result in unresolved double
tracks which deposit in both SDDs, on average, twice the energy loss of
one particle. A two-dimensional cut on both dE/dx measurements strongly
rejects such very close pairs without impairing reconstruction
efficiency by more than a few percent.  (ii) For opening angles larger
than 5 mrad but below 12 mrad -- the double ring resolution of the RICH-1
-- most double tracks will be resolved in the SDDs, but they match to
only one ring in RICH-1. Imposing an upper limit on the number of
Cherenkov photons removes a large fraction of tracks corresponding to
these pairs.  (iii) Pairs with even larger opening angles are recognized
if the second Cherenkov ring in RICH-1 and a corresponding trajectory in
the SDD system are found.  Tracks are rejected if the neighboring ring
is closer than 50 mrad. In order to simplify the
comparison to the S-Au analysis, we require the pair opening angle to be
larger than 35 mrad.  The combined effect of all rejection cuts improves
the signal-to-background
ratio by more than one order of magnitude. \\

\begin{figure*} 
\epsfig{file=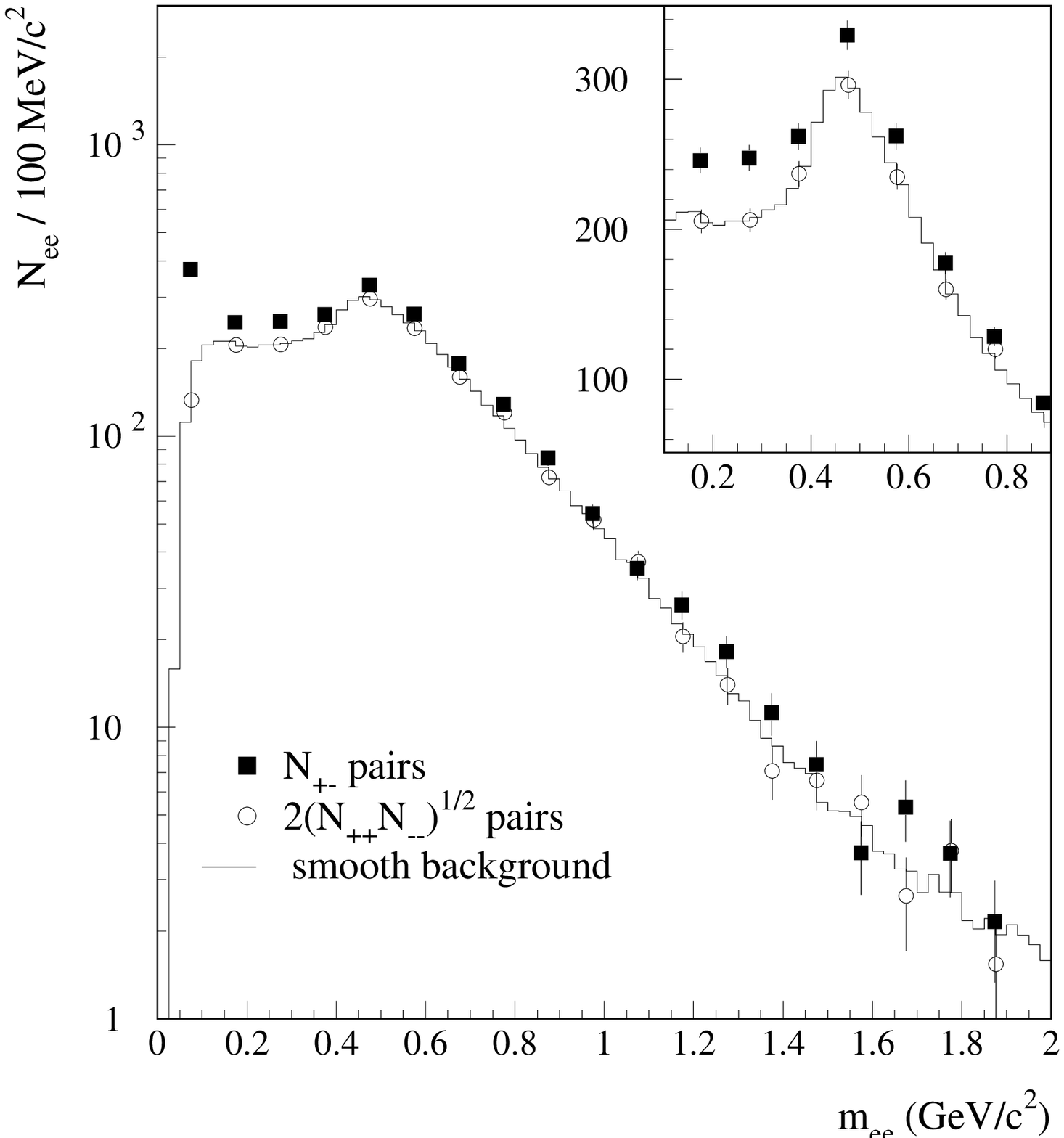,width=\textwidth}
\caption{{} Comparison of the invariant mass distributions of unlike-
  and like-sign pairs. The histogram gives the smoothed like-sign spectrum 
  which represents the best estimate of the combinatorial pair
  background (see text). The insert displays the low-mass region on a
  linear scale.}
\label{figure:raw}
\end{figure*}

The remaining combinatorial background in the
$\mathrm{e}^+\mathrm{e}^-$ sample is estimated by the number of like
sign pairs. In order to calculate the pair signal $S$, the like sign
contribution is subtracted from the $\mathrm{e}^+\mathrm{e}^-$-sample
such that $S = N_{+-}-2\sqrt{N_{++}N_{--}}$.  The mass distribution of
unlike-sign and of the geometrical mean of like-sign pairs is shown in
Fig.~\ref{figure:raw}.  In order to reduce statistical bin-to-bin
fluctuations, the like-sign spectra were smoothed before subtraction.
The smoothed pair background is created from tracks which are
generated randomly according to the measured $p_\perp$, $\theta$ and
$\phi$ distributions of tracks from like-sign pairs. By this
technique, an excellent description of the like-sign mass, $p_\perp$
and opening angle distributions is achieved in all observed phase
space selections.

The final sample contains 1038 $\pm$ 55 reconstructed
$\mathrm{e}^+\mathrm{e}^-$ pairs for invariant masses below 0.2
GeV/c$^2$ at a signal-to-background ratio of 1.05. For masses above 0.2
GeV/c$^2$ we find 5834 $\mathrm{e}^+\mathrm{e}^-$, 2494
$\mathrm{e}^+\mathrm{e}^+$, and 2696 $\mathrm{e}^-\mathrm{e}^-$ pairs
resulting in a net signal of $648 \pm 105$ pairs with a
signal-to-background ratio of 1/8. The difference spectrum turns out to
be stable in shape and yield with respect to details of the rejection
cuts, as elaborated below.

The resulting mass spectrum of pairs has to be corrected for efficiency.
The fraction of pairs which are reconstructed by our procedure is
estimated using a Monte-Carlo simulation in which individual lepton
pairs from neutral meson decays are generated (see below for more
details). The full detector response is modeled with a GEANT-based
\cite{geant} simulation and the specific detector properties.  Lacking
precise knowledge of how to generate complete events -- in particular
beam environment background -- these simulated pairs are embedded into
real data and are then analyzed using the standard software chain.  On
average about 50\% of the tracks are reconstructed, implying a
pair reconstruction efficiency of about 25\%.  The background rejection
cuts discussed above, however, reduce the number of reconstructed pairs
to about 12\%.  The reconstruction efficiency also depends on the
multiplicity in the event. It decreases considerably with increasing
multiplicity from 16\% for dN$_{ch}$/d$\eta$=120 to 9\% for
dN$_{ch}$/d$\eta$=350. No significant mass dependence of the efficiency
is observed.  

\begin{figure*}
\epsfig{file=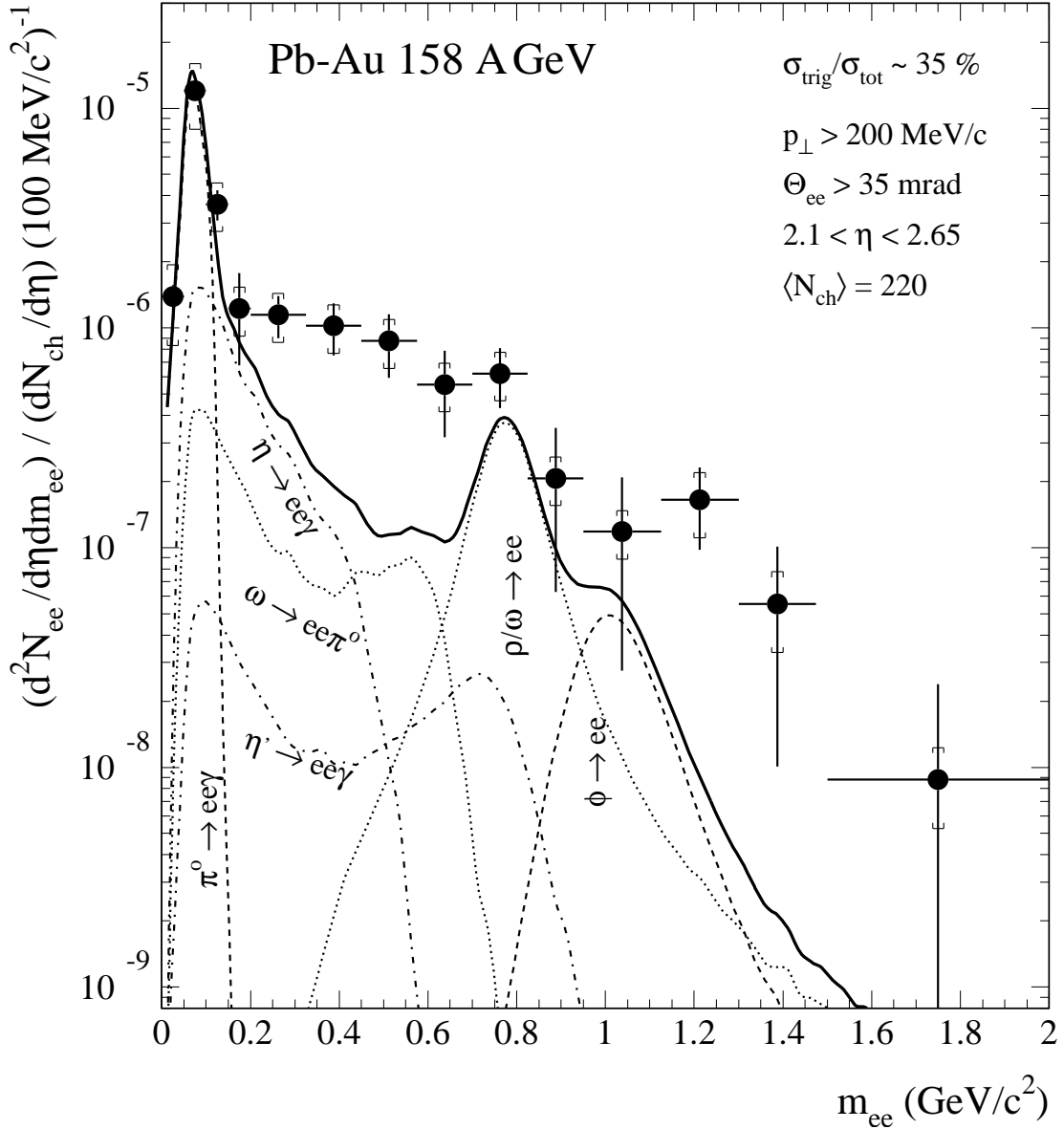, width=\textwidth}
\caption{{} Inclusive invariant $\mathrm{e}^+\mathrm{e}^-$ mass spectrum
  in 158 A~GeV Pb-Au collisions normalized to the observed charged-particle
  density. The statistical errors of the data are shown
  as bars, the systematic errors are given independently as brackets.
  The full line represents the $\mathrm{e}^+\mathrm{e}^-$ yield from
  hadron decays scaled from p-induced collisions. The contributions of
  individual decay channels are also shown.  }
  \label{figure:dndm}
\end{figure*}

Fig.~\ref{figure:dndm} shows the resulting invariant-mass spectrum of
$\mathrm{e}^+\mathrm{e}^-$ pairs.  The event sample corresponds to the
most central 35\% of the geometrical cross section with an average
charged multiplicity of $\langle$dN$_{ch}$/d$\eta\rangle$ = 220 in our
acceptance\footnote{ In an earlier
  presentation of a preliminary analysis the charged-particle density
  used was higher by 15\% due to an overestimate of a pileup correction
  \cite{ceres:qm96}.}.  The pair density has been normalized to
$\langle$dN$_{ch}$/d$\eta\rangle$ measured in the pseudorapidity region
2.1 to 2.65 by the silicon drift detectors. 
The statistical errors of the data are shown as
bars whereas systematic errors are given separately as brackets.  The
systematic errors contain contributions from the background smoothing
procedure, the uncertainties of the absolute background level, the
reconstruction efficiency and the determination of the average
multiplicity.  They add to 25-40\%, depending on mass, and are strongly
correlated for different mass bins.

Due to the small signal-to-background ratio, the background
subtraction is of crucial importance for the reliability of the
signal. In our case, this is even more so, because within the present
statistics the spectral shape of signal and background are not very
different for masses above 200 MeV, thus raising the question whether
the combinatorial background has been subtracted correctly.  For each
analysis cut, background rejection power and pair reconstruction
efficiency have been optimized simultaneously for four bins of
charged-particle multiplicity. The efficiency is obtained from the
Monte-Carlo simulation, while the rejection power is judged from the
measured like-sign pair background. The size of the signal itself was
disregarded in the optimization procedure in order to avoid the risk of
being mislead by statistical fluctuations in the data.  We have
performed two checks on the stability of our results:  (i)~All
background rejection cuts have been simultaneously varied, without
bias and at random within windows of $\approx\pm$15\% around the
optimum values. Several hundred cut settings were investigated. After
efficiency correction the result remained within the systematic
errors quoted. (ii)~The analysis was repeated using different
combinations of cuts, once without applying the final background
rejection cuts and in a second test with very much relaxed 
quality cuts in the tracking but including the final rejection cuts.
The signal-to-background ratio deteriorated from 1/8 to
1/25 and 1/30, respectively. Although the background level is
larger by a factor of more than $\sim$3, the final result remained
stable within statistical errors. 

The data are compared to the expected $\mathrm{e}^+\mathrm{e}^-$ pair
yield from decays of $\pi^0$, $\eta$, $\eta^\prime$, $\rho$, $\omega$,
and $\phi$, in line with ref.~\cite{ceres:pbegamma,ceres:pbe} where a more
detailed description can be found. The relative meson abundances are
assumed to be independent of the collision system. Their yield is scaled
proportional to $\langle$dN$_{ch}$/dy$\rangle$. The rapidity
distribution for pions is deduced from a Gaussian fit to measured data
\cite{y+pt:na49}, for heavier mesons the width of the rapidity
distribution is scaled using the maximum possible rapidity for a given
particle species of mass $m$ as $y_{max}(m)/y_{max}(\pi)$.  Similarly,
the pion $p_\perp$ distribution is generated according to a fit of
data from various experiments \cite{y+pt:na49,pt:na44,pt:wa98}.
Higher mass mesons are modeled according to measured kaon $p_\perp$
distributions, assuming $m_\perp$ scaling \footnote{ Due to our
relatively low $p_\perp$ cut of 200 MeV/c only the contribution of
$\pi^\circ$ and $\eta$ is influenced by the assumed $p_\perp$
distribution. Tracks from higher-mass mesons pass the $p_\perp$ cut
rather independent of the assumed distribution. Therefore, including for
example an increase of the slope parameter with particle mass as
observed in central Pb-induced collisions \cite{y+pt:na49,pt:na44},
would not alter our prediction of the e$^+$e$^-$ contribution from
meson decays.}.  Finally, the laboratory momenta of dileptons produced
in decays of these mesons are convoluted with the experimental
resolution and acceptance. The generated events were analyzed using
the same $p_\perp$ and opening-angle cuts as in the data analysis.

The invariant-mass spectrum from meson decays is in good agreement
with data below 200~MeV/c$^2$ where the expected contribution is
dominated by $\pi^0$ Dalitz decays. Above 200~MeV/c$^2$, the measured
pair yield is clearly enhanced, most pronounced in the region from 300
to 700~MeV/c$^2$. Integrating over this mass range, the
combined yield of the `hadronic cocktail' is exceeded by a factor of
5.8 $\pm$ 0.8(stat) $\pm$ 1.5 (syst). The enhancement, however,
extends up to the highest observed masses. In the larger interval
0.2 $\leq$ m $\leq$ 2.0 GeV/c$^2$, the enhancement factor is still 3.5
$\pm$ 0.4 (stat) $\pm$ 0.9 (syst).  The shape expected from the
generator spectrum has little correspondence to the data. In
particular, the resonance structures of $\rho/\omega$ and $\phi$ are
not visible albeit indicated by a sharply dropping yield on the
high-mass side.  The detection of the $\rho/\omega$ peak clearly is
within the capabilities of the mass resolution of
$\sigma_m/m_{\rho/\omega} \sim 8\%$, as can be seen from the solid
line in Fig.~\ref{figure:dndm}. In contrast, the peak was observed
as a prominent feature in the invariant-mass spectra of
$\mathrm{e}^+\mathrm{e}^-$ pairs from p--Be and p--Au collisions, with 
a mass resolution of only 11\% at m$_{\rho/\omega}$
\cite{ceres:pbe}.

The large impact parameter range covered by the data allows for a study
of the multiplicity dependence of the pair yield. If all pairs were
originating from decays of produced hadrons as scaled up from p--A
collisions, the pair yield should be proportional to charged-particle
multiplicity, as does the yield of produced hadrons in the generator.
Compared to this well-defined reference, any stronger-than linear
dependence of the $\mathrm{e}^+\mathrm{e}^-$ yield on N$_{ch}$ indicates
that pairs evolve from higher generations of hadrons or partons formed
inside the firecylinder, by any thermal hadro-chemical reaction in
general, or pion annihilation $\pi\pi\rightarrow\rho$ followed by
virtual photon radiation, in particular.

\begin{figure}
\epsfig{file=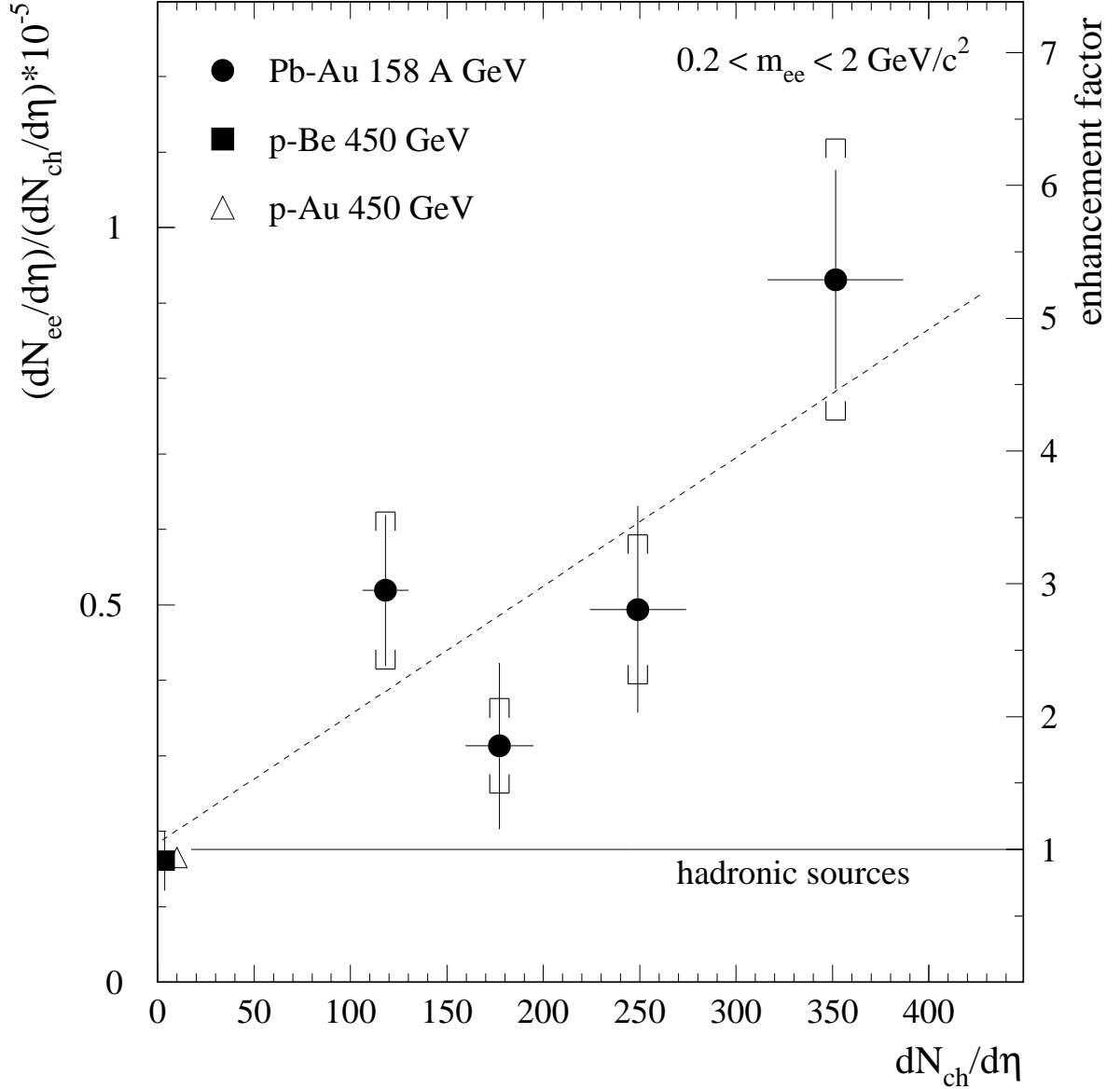, width=\textwidth}
\caption{{} Multiplicity dependence of the normalized pair yield for
  Pb-Au collisions. The error on the average multiplicity of each sample
  is given as horizontal bar. Statistical and systematic uncertainties
  of the pair yield are shown separately. The dashed line represents a
  fit to the Pb-Au data assuming a quadratic multiplicity dependence.
  The solid line shows the expectation from hadronic sources which
  corresponds to the integral of the contributions shown in
  Fig.~\ref{figure:dndm}.  The scale on the right hand side quantifies
  the enhancement above the expectation from hadron sources.  Also shown
  is the normalized yield measured in p-nucleus collisions {\protect
    \cite{ceres:pbe} }.  }
\label{figure:dndnch}
\end{figure}

The sample is divided into four subsets in charged-particle
multiplicity containing about equal numbers of
events. Fig.~\ref{figure:dndnch} gives the electron pair density
normalized to charged-particle density integrated over the mass range
0.2 $\leq$ m $\leq$ 2 GeV/c$^2$ for each of the multiplicity bins.
Also shown are the CERES results for the lighter collision systems
p-Be and p-Au \cite{ceres:pbe}. The reference assumption is
illustrated by the horizontal line in the figure, the
normalized pair density would remain constant. A significant
rise of the normalized pair yield is observed, such that the
dilepton yield grows faster than linear with multiplicity. This is
illustrated by the dashed line which shows a linear fit of the Pb-Au
data. With an error of $\sim 0.2$ the perfect agreement of the intercept
at $\langle$dN$_{ch}$/dy$\rangle$=0 with the yield observed in 
p-induced collisions and the expectation from hadron decays seems 
fortuitous. By applying a non-parametric statistical test, the
hypothesis of a linear scaling of the dilepton yield from p-Be to
central Pb-Au can be excluded with a 94\% confidence level.

\begin{figure}  
\epsfig{file=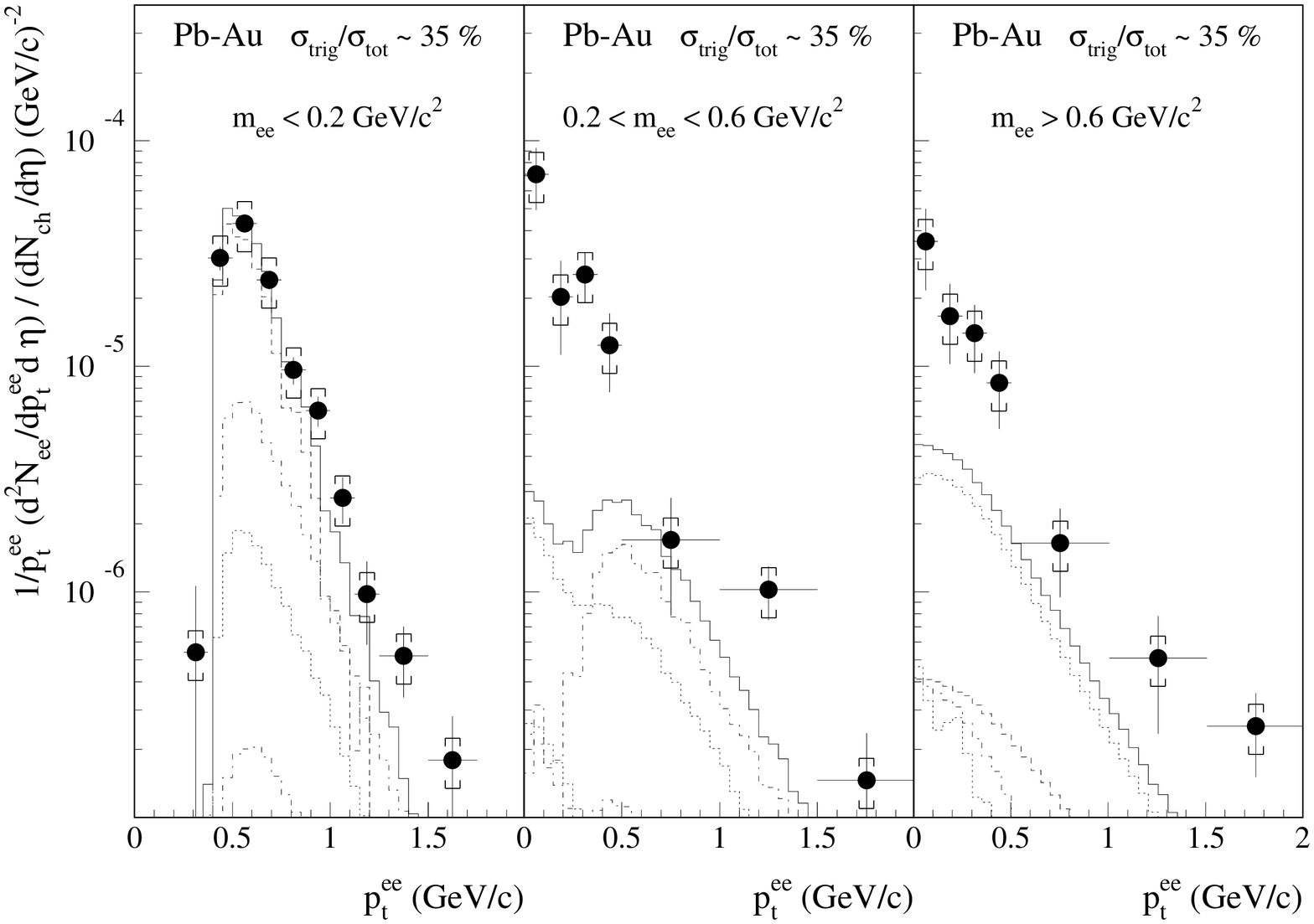, width=\textwidth}
\caption{{} Inclusive $\mathrm{e}^+\mathrm{e}^-$ pair $p_\perp$ spectra 
  in 158 GeV/nucleon Pb-Au collisions for three different pair mass
  ranges. Statistical and systematic errors are given separately. The
  histograms give the $p_\perp$-distributions expected from individual pardon
  decay channels and their sum. The normalization of the hadron decay
  contribution is identical to the one used in Fig.2.
  }
\label{figure:pairpt}
\end{figure}

The statistics of the present data sample is sufficient to extract
pair transverse-momentum spectra for three invariant-mass regions.  The
experimental data are plotted in Fig.~\ref{figure:pairpt}. One should
note that the single-track $p_\perp$-cut strongly affects the $p_\perp$
distribution for masses below 400 MeV/c$^2$, dominating the shape of the
pair $p_\perp$ spectra at low $p_\perp$. The generator curves shown are
folded with momentum resolution and normalized as was discussed
above. For masses below 200 MeV/c$^2$ the $p_\perp$ distribution agrees
with the prediction for $\pi^0$ Dalitz decays (left panel). For higher
masses, the enhancement is visible over the entire $p_\perp$ range but
it significantly increases towards very low $p_\perp$ 
for m$\geq$ 200 MeV/c$^2$.

In summary, we have presented first results of
$\mathrm{e}^+\mathrm{e}^-$ pair production in Pb-Au collisions. Above
masses of 0.2 GeV/c$^2$ the continuum is enhanced compared to the
yield expected from neutral meson decays by scaling from
p-induced reactions.  This result corroborates previous findings with
sulfur beam where a similar enhancement of the continuum was found for
$\mu^+\mu^-$ \cite{helios3:sw} and $\mathrm{e}^+\mathrm{e}^-$ 
\cite{ceres:sau} pairs. For central Pb-Au and S-Au collisions the spectral
shape and magnitude of the $\mathrm{e}^+\mathrm{e}^-$ pair enhancement
are similar. The larger statistics available for
Pb-Au collisions allowed to extend the mass range up to 2
GeV/c$^2$. We have found evidence for a strong increase of the
pair excess with centrality as expected for thermal
radiation. First data on $p_\perp$-spectra show that the yield is
amplified over the entire range, but it is strongest at the
lowest $p_\perp$ measured.\\

The CERES collaboration acknowledges the good performance of the CERN
PS and SPS accelerators. We are grateful for support by the German
BMBF, the U.S.~DoE, the MINERVA Foundation, 
the German-Israeli Foundation for Scientific Research and Development,
and the Israeli Science Foundation.


\newpage

\newpage
 
\newpage

\vspace*{-2cm}

\newpage


\begin{references}

\bibitem{helios3:sw} M.~Masera~{et~al.} (HELIOS/3-Collaboration), 
                     Nucl.~Phys.~{A590} (1995) 93c.
\bibitem{na38:su}    C.~Louren\c{c}o,
                     Doctoral Thesis, Universidade Tecnica de Lisboa (1995). 
\bibitem{ceres:sau}  G.~Agakichiev~{et~al.} (CERES-Collaboration),
                     Phys.~Rev.~Lett.~{75} 1272 (1995).
\bibitem {ref1}      I.~Tserruya,  Nucl.~Phys.~{A590} (1995) 127c.

\bibitem {ref2}      A.~Drees, Nucl.~Phys.~{A610} (1996) 536c;
                     Nucl.~Phys.~{A630} (1998) 449c.
\bibitem{ceres:pbegamma}  G.~Agakichiev~{et~al.} (CERES-Collaboration), 
                     Neutral meson production in p-Be and p-Au
                     collisions at 450 GeV beam energy, 
                     Z.~Phys.~C.~(in print).
\bibitem{ceres:pbe}  G.~Agakichiev~{et~al.} (CERES-Collaboration), 
                     Systematic study of low-mass electron pair
                     production in p-Be and p-Au collisions at 450 GeV/c, 
                     Z.~Phys.~C.~(in print).
\bibitem {li-ko-brown} G.\thinspace Q.~Li, C.\thinspace M.~Ko, and G.~Brown, 
                     Phys.~Rev.~Lett.~{75} (1995) 4007;
                     Nucl.~Phys.~{A606} (1996) 568.
\bibitem {cassing}   W.~Cassing, W.~Ehehalt, and C.\thinspace M.~Ko, 
                     Phys.~Lett.~{B363} (1995) 35;
                     W.~Cassing, W.~Ehehalt, and I.Kr\'alik, 
                     Phys.~Lett.~{B377} (1996) 5.
\bibitem{wambach}    R.~Rapp, G.~Chanfray, and J.~Wambach, 
                     Phys.~Rev.~Lett.~{76} (1996) 368;
                     Nucl. Phys.~{A617} (1997) 472.
\bibitem{theorylist} D.\thinspace K.~Srivastava, B.~Sinha, and C.~Gale, 
                     Phys.~Rev.~{C53} (1996) R567;
                     J.~Sollfrank~{et~al.}, 
                     Phys.~Rev.~{C55} (1997) 392;
                     C.\thinspace M.~Hung and E.\thinspace V.~Shuryak, 
                     Phys.~Rev.~{  C56} (1997) 453;
                     V.~Koch and C.~Song, 
                     Phys.~Rev.~{C54} (1996) 1903;
                     R.~Baier, M.~Dirks, and K.~Redlich,
                     Phys.~Rev.~{D55} (1996) 4344;
                     J.~Murray, W.~Bauer, and K.~Haglin, hep-ph/9611328,
                     to be published;
                     L.~Winckelmann~{et~al.}, 
                     Nucl.~Phys.~{A610} (1996) 116c.
\bibitem {brown-rho} G.E.~Brown and M.~Rho, 
                     Phys.~Rev.~Lett.~{66} (1991) 2720.
\bibitem {hat-lee}   T.~Hatsuda and S.\thinspace H.~Lee, 
                     Phys.~Rev.~{C46} (1992) R34.
\bibitem{ceres:qm96} T.~Ullrich~{et~al.} (CERES-Collaboration), 
                     Nucl.~Phys.~{A610} (1996) 317c.
\bibitem{ceres:bari} R.~Baur~{et~al.} (CERES-Collaboration),
                     Nucl.~Inst.~Meth.~{A343} (1994) 87.
\bibitem{geant}      R.~Brun {et~al.},
                     CERN DD/EE/84-1 (1984), unpublished.
\bibitem{y+pt:na49}  S.\thinspace V.~Afanasiev~{et~al.} (NA49-Collaboration), 
                     Nucl.~Phys.~{A610} (1996) 188c.
\bibitem{pt:na44}    I.\thinspace G.~Bearden~{et~al.} (NA44-Collaboration), 
                     Nucl.~Phys.~{A610} (1996) 175c.
\bibitem{pt:wa98}    M.~Aggarwal~{et~al.} (WA98-Collaboration), 
                     Nucl.~Phys.~{A610} (1996) 200c.
\end{references}
\end{document}